\documentclass[12pt]{article}
\usepackage{cite,epsfig,amssymb,amsmath,graphicx,color}
\newcommand{\be}{\begin{equation}}
\newcommand{\ee}{\end{equation}}
\newcommand{\bear}{\begin{eqnarray}}
\newcommand{\eear}{\end{eqnarray}}
\newcommand{\ba}{\begin{array}}
\newcommand{\ea}{\end{array}}
\newcommand{\nn}{\nonumber}

\begin{document}

\begin{center}
{{{\Large \bf Generalized Lee-Wick Formulation
\\ from Higher Derivative Field Theories}
}\\[15mm]
{Inyong Cho}
\\[2mm]
{\it School of Liberal Arts, Seoul National University of Technology,
Seoul 139-743, Korea}\\
{\tt iycho@snut.ac.kr}\\[5mm]
{O-Kab Kwon}\\[2mm]
{\it Department of Physics, BK21 Physics Research Division,
Institute of Basic Science,\\
Sungkyunkwan University, Suwon 440-746, Korea}\\
{\tt okab@skku.edu} }
\end{center}

\vspace{10mm}

\begin{abstract}

We study a higher derivative (HD) field theory with an arbitrary order
of derivative for a real scalar field. The degree of freedom for
the HD field can be converted to multiple fields with canonical
kinetic terms up to the overall sign. The Lagrangian describing the
dynamics of the multiple fields is known as the Lee-Wick (LW) form.
The first step to obtain the LW form for a given HD Lagrangian is to
find an auxiliary field (AF) Lagrangian which is equivalent to the
original HD Lagrangian up to the quantum level.
Till now, the AF Lagrangian has been studied only for $N=2$ and $3$
cases, where $N$ is the number of poles of
the two-point function of the HD scalar field. We construct the AF
Lagrangian for arbitrary $N$. By the linear combinations of AF
fields, we also obtain the corresponding LW form. We find the
explicit mapping matrices among the HD fields, the AF fields, and the
LW fields. As an exercise of our construction, we calculate the
relations among parameters and mapping matrices for $N=2,3$, and 4 cases.
\end{abstract}
\newpage

\section{Introduction}

Lee and Wick (LW) constructed quantum electrodynamics with the
higher derivative (HD) propagator for a photon~\cite{Lee:1969fy}.
The HD term in the denominator of the propagator improves
the ultraviolet convergence of the Feynman diagram since the propagator
falls off more quickly as momentum grows. A minimal set of the HD term
leads to an additional physical pole in the propagator, corresponding
to a massive LW-photon with the wrong-sign residue.
It seems that the wrong sign gives rise to the instability or the violation of
unitarity in the theory. Lee and Wick proposed a deformation of integration
contours in the Feynman diagram so that the theory can be free from the instability
and the violation of unitarity.

Based on the idea of Lee and Wick~\cite{Lee:1969fy},
Lee-Wick Standard Model (LWSM) has been proposed as a candidate
to solve the hierarchy problem in the Standard Model~\cite{Grinstein:2007mp}.
Every field in the LWSM has a higher derivative kinetic term.
Using the auxiliary field (AF) method,
the HD term can be converted into the degree of freedom of a massive field
with wrong-sign kinetic term, referred as the LW partner. Since the
LW partner can decay into ordinary fields, the `wrong sign' does not
cause the violation of unitarity at macroscopic
scales~\cite{Lee:1969fy,Grinstein:2007iz,Grinstein:2008bg}.
Due to the presence of the LW partner, the radiative quantum correction
to the Higgs-boson mass-squared is free from the quadratic divergence,
which can be a resolution of the hierarchy problem in the Standard Model.
The simple mechanism solving the hierarchy problem has motivated
phenomenological studies~\cite{Rizzo:2007ae,Espinosa:2007ny,
Dulaney:2007dx,Krauss:2007bz,Grinstein:2008qq,Alvarez:2008za,Underwood:2008cr,
Carone:2008bs,Fornal:2009xc,Rodigast:2009en,Carone:2009nu,Alvarez:2009af,
Carone:2008iw,Carone:2009it}.
The LWSM was also extended to studying cosmology~\cite{Li:2005fm}.

In LWSM~\cite{Grinstein:2007mp}, a minimal set of the HD term was considered.
The number of poles in the two-point function of each field in LWSM is two,
representing the number of physical degrees of freedom which correspond to
the ordinary Standard Model particle and its LW partner. From now on, we refer
the number of poles in the two-point function of the HD theory as $N$.
That is, LWSM~\cite{Grinstein:2007mp} is an $N=2$ HD theory. In general we can
include the HD terms beyond the minimal set. In this direction, Carone and
Lebed in Ref.~\cite{Carone:2008iw} constructed the $N=3$ HD theory by
providing the mapping among the HD Lagrangian, the AF Lagrangian,
and the LW form. Here the LW form represents
the Lagrangian having canonical quadratic terms aside from the overall sign.
While the $N=2$ HD theory has a single LW partner with wrong-sign quadratic
terms in the LW form, the $N=3$ HD theory has two LW partners
one of which has the correct sign (corresponding the
ordinary particle) while the other has the wrong sign.
That is, there is an additional ordinary
LW partner in the LW form of the $N=3$ HD theory.

Due to the existence of the additional ordinary LW partner,
the $N=3$ HD theory is qualitatively different from the $N=2$ HD theory.
For instance, the gauge coupling unification in the Standard Model,
which is difficult to be realized in the $N=2$ HD theory, can be achieved
at the one-loop level in the $N=3$ HD theory without introducing
additional fields~\cite{Carone:2009it}. Therefore, the investigation of
the HD theory beyond the minimal set is an interesting subject.

With this motivation, we generalize the Carone and Lebed's
construction~\cite{Carone:2008iw} for a scalar field.
We consider a general HD Lagrangian of a self-interacting real scalar
field~\footnote{We can rewrite the kinetic term as
the canonical form, such as $\frac{1}{2}\phi\Box\phi = -\frac{1}{2}
\nabla_\mu\phi\nabla^\mu\phi$, by neglecting the total derivative
term. However, mostly we use D'Alembertian $\Box\equiv \frac{1}{\sqrt{-g}}
\partial_\mu\big(\sqrt{-g}g^{\mu\nu}\partial_\nu\big)$, instead of $\nabla_\mu$
for simplicity.},
\begin{align}\label{HDL}
{\cal L}_{{\rm HD}}^{(N)}
=\frac{1}{2}\sum_{n=1}^N (-1)^{n+1}a_n\phi \Box^n\phi
-\frac{1}{2} m^2\phi^2+ {\cal L}_{{\rm int}}(\phi)
\end{align}
with $a_1=1$, where $N$ indicates the number of physical poles
in the propagator of $\phi$, $a_n$ is the coefficient of mass dimension
$[a_n] = 2-2n$, and the last term represents interactions.
The equation of motion for $\phi$ is given  by
\begin{align}\label{HD-eom}
\sum_{n=1}^N (-1)^{n+1} a_n\Box^n\phi - m^2\phi
+ {\cal L}_{{\rm int}}^{'} = 0,
\end{align}
where ${\cal L}_{{\rm int}}^{'}= d{\cal L}_{\rm int}/d\phi$.
We construct the AF Lagrangians for $N=$ even and $N=$ odd cases
separately for the HD Lagrangian.
By integrating out the auxiliary fields, one can
restore the original HD Lagrangian (\ref{HDL}) and the equation of motion
(\ref{HD-eom}).
Since the auxiliary fields are linear or quadratic in the AF Lagrangian
(see section 2), the HD Lagrangian and the corresponding AF Lagrangian
are equivalent at the quantum level also.
We obtain the explicit transformation from the HD field to the AF
field.\footnote{In this work, the ``AF field'' denotes
the auxiliary field plus the ordinary field in the AF Lagrangian.}

For general $N$, we construct the transformation from the HD field
to the LW field. We also show that this LW field can be transformed to
the AF field.
Therefore, the AF Lagrangian can be written
in the LW form by linear combinations of the AF fields,
which indicates that the HD Lagrangian and the LW form are also
equivalent at the quantum level.\footnote{For a pure scalar-field theory,
one can show directly that the HD lagrangian is equivalent to the LW form
at the quantum level, without the aid of the AF lagrangian.
However, introducing the AF lagrangian shall be useful
for future investigation involving gauge fields.}

Studying the HD field theory has a long history.
In 1950, Pais and Uhlenbeck~\cite{Pais:1950za} investigated
the HD theories in terms of the low-order theories
(so, equivalently the LW form)
in order to study purely quantum-mechanical theories.
In the modern point of view, we can also find the origin of
the HD field theory from string theory,
for example,  the tachyon effective action of the truncated
open string field theory, the $p$-adic string theory, and etc.
One of the interesting features of these theories is
the presence of the infinite order of derivatives. Recently there
has been much interest in applying these nonlocal theories with infinite
order of derivatives to cosmological
models~\cite{Aref'eva:2005gg,Aref'eva:2006et,Barnaby:2006hi,Lidsey:2007wa,
Barnaby:2007yb,Calcagni:2007ru,Joukovskaya:2007nq,Koivisto:2008xfa,
Koshelev:2009ty,Calcagni:2009dg,Vernov:2009vf}
and to analyzing the mathematical structure of those
theories~\cite{Barnaby:2007ve,Calcagni:2007ef,Mulryne:2008iq,Barnaby:2008tc}.
The supersymmetric extension of the HD field theories is also an interesting
subject~\cite{Buchbinder:1996je}.
Our result can be accommodated to this type of interesting work.

This paper is organized as follows.
In section 2, we construct the AF Lagrangian from the HD Lagrangian
for the $N=$ even and the $N=$ odd cases separately.
In section 3, we construct the LW form from the HD Lagrangian,
and obtain an explicit transformation from the LW field to the HD field.
We also find a transformation from the LW field to the AF field.
We apply our formalism for $N=2,3$, and $4$ cases, and provide the results.
We conclude in section 4.

\section{Generalized Auxiliary Field Lagrangian}\label{sec-AFL}
By introducing auxiliary fields we can write an AF Lagrangian.
Integrating out these auxiliary fields from the Lagrangian reproduces
the HD Lagrangian.
The AF Lagrangian has the ordinary kinetic terms.
The number of the fields including those auxiliary fields is $N$ .
Until now, the AF Lagrangians for
$N=2$~\cite{Grinstein:2007mp} and $N=3$~\cite{Carone:2008iw} cases have been
constructed. In this work, we construct an AF Lagrangian
for general $N$. We  shall separately treat the $N=2\hat N$
and $N=2\hat N+1$ cases with positive integer $\hat N$.

\subsection{$N=2\hat N$ case}\label{Evencase}

In this subsection we construct a generalized AF Lagrangian
which is equivalent to the original HD Lagrangian (\ref{HDL}) for
arbitrary $N=2\hat N$, and obtain the transformation between the AF and
the HD fields.
The form of the Lagrangian is given by
\begin{align}\label{AFL-even}
{\cal L}_{{\rm AF}}^{(2\hat N)} =
\frac12\sum_{n=1}^{\hat N}\varphi_n(\Box-Q_n)
\varphi_n + \sum_{n=1}^{\hat N-1}\chi_n(\Box\varphi_n -R_n\varphi_{n+1})
+\chi_{\hat N}\Box\varphi_{\hat N} +\frac{R_{{\hat N}}^2}{2Q_{\hat N+1}}
\chi_{\hat N}^2 + {\cal L}_{{\rm int}}(\varphi_1),
\end{align}
where $\varphi_n$ and $\chi_n$ are real
scalar fields and $R_n$ and $Q_n$
are constant parameters of mass dimension 2.\footnote{We
introduced $R_{\hat N}$ in (\ref{AFL-even}) for the formal
consistency, but we can treat $R_{\hat N}^2/Q_{\hat N+1}$
as a single parameter.}
Using the relations among parameters ($m$, $a_n$) in (\ref{HDL})
and ($Q_n$, $R_n$) in (\ref{AFL-even}),
we can verify that ${\cal L}_{{\rm HD}}^{(2\hat N)}$ and
${\cal L}_{{\rm AF}}^{(2\hat N)}$ are equivalent.

In order to verify that the AF Lagrangian (\ref{AFL-even})
is equivalent to the HD Lagrangian (\ref{HDL}) with even number of poles,
we first integrate out the auxiliary field $\chi_n$ in (\ref{AFL-even}).
In order to do this we write the equation of motion
for the auxiliary field $\chi_n$, and then
$\varphi_n$ ($n>1$) is expressed in terms of $\varphi_1$:
\begin{align}\label{AF-chi-even}
\chi_1 &:\,\varphi_2 = \frac{1}{R_1}\Box\varphi_1 = S_1\Box\varphi_1,
\nn \\
\chi_2 &:\, \varphi_3 = \frac{1}{R_2} \Box\varphi_2= S_2\Box^2\varphi_1,
\nn \\
&\hskip 2cm \vdots
\nn \\
\chi_{\hat N-1} &:\,\varphi_{\hat N} = \frac1{R_{\hat N-1}}
\Box\varphi_{\hat N-1} = S_{\hat N-1}\Box^{\hat N-1}\varphi_1,
\nn \\
\chi_{\hat N} &: \chi_{\hat N} =-\frac{Q_{\hat N+1}}{R_{{\hat N}}^2}\,
\Box\varphi_{\hat N} = -\frac{S_{\hat N}Q_{\hat N+1}}{R_{{\hat N}}}\,
\Box^{\hat N}\varphi_1,
\end{align}
where we defined
\begin{align}
S_n\equiv \prod_{i=1}^{n}\frac1{R_i}. \nn
\end{align}
Therefore, the auxiliary field $\varphi_n$ is expressed by
the HD field $\phi$,
\be
\varphi_n = S_{n-1}\phi_{n-1},
\qquad (n=2,...,\hat N),
\label{phi-to-variphi}
\ee
where we defined $\phi_n \equiv \Box^n \phi$, and set $\phi=\varphi_1$.

Since the auxiliary field $\chi_n$ is linear or quadratic in the Lagrangian
(\ref{AFL-even}), the equations of motion (\ref{AF-chi-even})
obtained by the variation of $\chi_n$ are exact at the quantum level also.
Plugging the equations in (\ref{AF-chi-even}) into
(\ref{AFL-even}), we obtain
\begin{align}\label{AFL-even2}
{\cal L}_{{\rm AF}}^{(2\hat N)}&=\frac12\sum_{n=1}^{\hat N}
S_{n-1}^2\varphi_1(\Box - Q_n)\Box^{2n-2}\varphi_1
-\frac12 S_{\hat N}^2Q_{\hat N+1} \varphi_1\Box^{2\hat N}\varphi_1
+{\cal L}_{{\rm int}}(\varphi_1)
\nn \\
&=\frac12\sum_{n=1}^{\hat N}S_{n-1}^2\phi\Box^{2n-1}\phi
-\frac12\sum_{n=1}^{\hat N+1}S_{n-1}^2Q_n\phi\Box^{2n-2}\phi
+{\cal L}_{{\rm int}}(\phi).
\end{align}

Next, we read the equation of motion for $\varphi_n$:
\begin{align}\label{AF-varphi-even}
\varphi_1 &:\, (\Box-Q_1)\varphi_1 + \Box\chi_1
+ {\cal L}_{{\rm int}}^{'}(\varphi_1) = 0,
\nn \\
\varphi_2 &:\,(\Box-Q_2)\varphi_2 + \Box\chi_2 = R_1\chi_1,
\nn \\
\varphi_3 &:\,(\Box-Q_3)\varphi_3 + \Box\chi_3 = R_2\chi_2,
\nn \\
&\hskip 1.5cm \vdots
\nn \\
\varphi_{\hat N} &:\,(\Box-Q_{\hat N})\varphi_{\hat N}
+\Box\chi_{\hat N}= R_{\hat N-1}\chi_{\hat N-1}.
\end{align}
Combining equations (\ref{AF-chi-even}) and (\ref{AF-varphi-even}),
we can express $\chi_n$ in terms of $\varphi_1(=\phi)$,
\begin{eqnarray}\label{chi-even}
\chi_n &=& \sum_{l=n}^{\hat N-1}\frac{S_l^2}{S_{n-1}}\,\Box^{2l-n+1}\varphi_1
-\sum_{l=n}^{\hat N}\frac{S_l^2 Q_{l+1}}{S_{n-1}}\,\Box^{2l-n}\varphi_1,
\qquad (n=1,...,\hat N), \\
&=& \left[ \sum_{l=n}^{\hat N-1} \frac{S_l^2}{S_{n-1}}
\left( \Box - Q_{l+1} \right) \Box^{2l-n}
- \frac{S_{\hat N}^2 Q_{\hat N+1}}{S_{n-1}}\,\Box^{2\hat N-n} \right] \phi,
\label{phi-to-chi}
\end{eqnarray}
where $S_0=1$. See Appendix A for the detail.
The auxiliary field $\chi_n$ is now expressed by the HD field $\phi$
via Eq.~\eqref{phi-to-chi}.

In order to get the relations between the HD parameter $a_n$ and
the AF parameters $Q_n$ and $S_n$,
let us compare ${\cal L}_{{\rm HD}}^{(2\hat N)}$ in (\ref{HDL}) and
${\cal L}_{{\rm AF}}^{2\hat N}$ in (\ref{AFL-even}).
We decompose
${\cal L}_{{\rm HD}}^{(2\hat N)}$ as
\begin{align}\label{HDL-even}
{\cal L}_{{\rm HD}}^{(2\hat N)} =
\frac12\sum_{n=1}^{\hat N} a_{2n-1}\phi\Box^{2n-1}\phi
-\frac12\sum_{n=1}^{\hat N+1} a_{2n-2} \phi\Box^{2n-2}\phi
+ {\cal L}_{{\rm int}}(\phi).
\end{align}
Comparing this equation and (\ref{AFL-even2}), we obtain the relations,
\begin{align}\label{rel-even}
&Q_1 = m^2, \quad Q_{n+1} = \frac{a_{2n}}{a_{2n+1}},
\quad \frac{Q_{\hat N+1}}{R_{{\hat N}}^2}\,S_{{\hat N}-1}^2 = a_{2\hat N},
\nn \\
&S_0^2 = 1,\quad S_n^2 = a_{2n+1}, \quad
(n=1,...,\hat N-1).
\end{align}
Then ${\cal L}_{{\rm AF}}^{2\hat N}$ and
${\cal L}_{{\rm HD}}^{2\hat N}$ are equivalent.
Since $Q_{\hat N+1}/R_{\hat N}^2$ is a single parameter
as we discussed before,
we can set $R_{\hat N}=1$ for computational convenience.
In this case, the parameter $Q_{\hat N+1}$ has mass dimension $-2$.

Now we can show that the HD field equation is reduced from the AF field
equation. Inserting the relation for $\chi_1$ in (\ref{chi-even}) into
the first equation in (\ref{AF-varphi-even}),
we can get the field equation for $\varphi_1$,
\begin{align}\label{AF-eom1}
(\Box - Q_1)\varphi_1 + \sum_{n=1}^{\hat N-1} S_n^2
(\Box-Q_{n+1}) \Box^{2n}\varphi_1
-S_{\hat N}^2Q_{\hat N+1}\Box^{2\hat N}\varphi_1
+ {\cal L}_{{\rm int}}^{'}(\varphi_1) = 0,
\end{align}
where $a_0=m^2$. Using the relation \eqref{rel-even} this equation
becomes the field equation \eqref{HD-eom} for the HD field
$\phi(=\varphi_1)$.
This completes the proof that the AF Lagrangian (\ref{AFL-even})
is equivalent to the HD Lagrangian (\ref{HDL})
with even number of physical poles.

Now we obtain the explicit transformation matrix between the HD field and
AF fields.
Substituting the relations in (\ref{rel-even}) into (\ref{phi-to-variphi})
and (\ref{phi-to-chi}), we can express the AF fields ($\varphi_n,\,\chi_n$)
as the HD field $\Box^n\phi$ in terms of the coefficient $a_n$,
\begin{align}\label{HDtoAF}
\varphi_n &= \sqrt{a_{2n-1}}\,\phi_{n-1},
\nn \\
\chi_n &=\sum_{l=n}^{\hat N-1}\frac{a_{2l+1}}{\sqrt{a_{2n-1}}}\,
\phi_{2l-n+1}-\sum_{l=n}^{\hat N}\frac{a_{2l}}{\sqrt{a_{2n-1}}}\,
\phi_{2l-n}
\nn \\
&=\sum_{m=n}^{2\hat N-n}(-1)^{m+n-1}\,\frac{ a_{m+n}}{\sqrt{a_{2n-1}}}\,
\phi_{m}\qquad (n=1,\cdots, \hat N).
\end{align}
Here, we assume that $S_n>0$.
The negative $S_n$ can also be chosen. (However,
the resulting AF Lagrangian has no difference from the positive $S_n$ case
since the signature change in $S_n$ means
the overall-signature change in the fields $\varphi_n$ and $\chi_n$.)
From the results in (\ref{HDtoAF}), we can read the transformation matrix
$U^{(e)}_{ij}$
satisfying
\begin{align}\label{HDtoAF3}
\xi_i = \sum_{j=1}^{2\hat N} U^{(e)}_{ij}\phi_{j-1},
\qquad (i,\,j=1,\cdots, 2\hat N),
\end{align}
where we combine ($\varphi_n,\,\chi_n$) to a column matrix $\xi$ with
$2\hat N$ components,
\begin{align}\label{defxie}
\xi_n \equiv \varphi_n,\qquad \xi_{\hat N+n}\equiv \chi_n, \qquad
(n=1,\cdots,\hat N).
\end{align}
The components of the transformation matrix $U^{(e)}_{ij}$ are given by
\begin{align}
U^{(e)}_{ij} =\left\{
\begin{array}{ll}
&\left.
\begin{array}{ll}
\sqrt{a_{2j-1}}\,\delta_{ij} \qquad\qquad &~
(1\le j\le\hat N)\\
0 \qquad\qquad & ~(\mbox{the others})
\end{array}
\hskip 3.17cm \right\} (1\le i \le\hat N) \\
&\left.
\begin{array}{ll}
(-1)^{i+j-\hat N}
\frac{a_{i+j-\hat N-1}}{\sqrt{a_{2i-2\hat N-1}}} &
(i-\hat N+1\le j\le 3\hat N-i+1) \\
0 & (\mbox{the others})
\end{array}
\right\} (\hat N+1\le i \le 2\hat N)
\end{array}\right..
\end{align}
We can also represent the HD fields $\phi_n$ in terms of the AF fields
by using inverse transformation of $U^{(e)}_{ij}$,
\begin{align}
\phi_{i-1} = \sum_{j=1}^{2\hat N} {U^{(e)}}^{-1}_{ij}\xi_j,
\end{align}
where ${U^{(e)}}^{-1}_{ij}(a_n)$ can be expressed as
${U^{(e)}}^{-1}_{ij}(Q_n,S_n)$ using~Eq.~\eqref{rel-even}.

As an example, we apply the AF Lagrangian (\ref{AFL-even}) for the $N=2$ case.
Then the AF Lagrangian (\ref{AFL-even}) is written as
\begin{align}\label{AFL-even3}
{\cal L}_{{\rm AF}}^{(2)} = \frac12\varphi_1(\Box-Q_1)
+ \chi_1\Box\varphi_1 + \frac1{2Q_2}\chi_1^2
+ {\cal L}_{{\rm int}}(\varphi_1),
\end{align}
where we set $R_2=1$ for simplicity.
With the identifications given in (\ref{rel-even}),
\begin{align}
\varphi_1=\phi,\quad Q_1 =m^2, \quad Q_2 = \frac{a_2}{a_1}=\frac1{M^2},
\end{align}
one can easily check that (\ref{AFL-even3}) is the well-known Lagrangian
for the $N=2$ HD Lagrangian~\cite{Grinstein:2007mp}.

\subsection{$ N=2\hat N +1$ case}\label{oddcase}
In this subsection we construct the AF Lagrangian with the odd number
of poles, and obtain the transformation between the AF and
the HD fields.
The Lagrangian can be written in the following way,
\begin{align}\label{AFL-odd}
{\cal L}_{{\rm AF}}^{(2\hat N+1)}
=\frac12\sum_{n=1}^{\hat N+1}\varphi_n(\Box-Q_n)
\varphi_n + \sum_{n=1}^{\hat N}\chi_n(\Box\varphi_n -R_n\varphi_{n+1})
+{\cal L}_{{\rm int}}(\varphi_1).
\end{align}
Using the similar method to the $N=2\hat N$ case in the previous subsection,
we verify that (\ref{AFL-odd}) is the AF Lagrangian 
of ${\cal L}_{{\rm HD}}^{(N)}$ in (\ref{HDL}) with the odd number of poles.
The equation of motion for the auxiliary field $\chi_n$ is given by
\begin{align}\label{AF-chi-odd}
\chi_1 &:\,\varphi_2 = \frac{1}{R_1}\Box\varphi_1 = S_1\Box\varphi_1,
\nn \\
\chi_2 &:\, \varphi_3 = \frac{1}{R_2} \Box\varphi_2= S_2\Box^2\varphi_1,
\nn \\
&\hskip 2cm \vdots
\nn \\
\chi_{\hat N} &:\,\varphi_{\hat N+1} = \frac1{R_{\hat N}}
\Box\varphi_{\hat N} = S_{\hat N}\Box^{\hat N}\varphi_1.
\end{align}
Therefore, in the same way with the previous section, we have
\be
\varphi_n = S_{n-1} \phi_{n-1},
\qquad (n=1,...,\hat N+1).
\label{phi-to-variphi2}
\ee
Plugging this equation into
${\cal L}_{{\rm AF}}^{(2\hat N+1)}$ in (\ref{AFL-odd}), we obtain
\begin{align}\label{AF-odd2}
{\cal L}_{{\rm AF}}^{(2\hat N+1)}&=\frac12\sum_{n=1}^{\hat N+1}
S_{n-1}^2\varphi_1(\Box - Q_n)\Box^{2n-2}\varphi_1 +
{\cal L}_{{\rm int}}(\varphi_1)
\nn \\
&=\frac12\sum_{n=1}^{\hat N+1} S_{n-1}^2\phi\Box^{2n-1}\phi
-\frac12\sum_{n=1}^{\hat N+1} S_{n-1}^2Q_n\phi\Box^{2n-2}\phi
+{\cal L}_{{\rm int}}(\phi).
\end{align}

Next, we write the equation of motion for $\varphi_n$:
\begin{align}\label{AF-varphi-eom}
\varphi_1 &:\, (\Box-Q_1)\varphi_1 + \Box\chi_1
+{\cal L}_{{\rm int}}^{'}(\varphi_1) = 0,
\nn \\
\varphi_2 &:\,(\Box-Q_2)\varphi_2 + \Box\chi_2 = R_1\chi_1,
\nn \\
\varphi_3 &:\,(\Box-Q_3)\varphi_3 + \Box\chi_3 = R_2\chi_2,
\nn \\
&\hskip 1.5cm \vdots
\nn \\
\varphi_{\hat N} &:\,(\Box-Q_{\hat N})\varphi_{\hat N} + \Box\chi_{\hat N}
= R_{\hat N-1}\chi_{\hat N-1},
\nn \\
\varphi_{\hat N+1} &:\,(\Box-Q_{\hat N+1})\varphi_{\hat N+1}
= R_{\hat N}\chi_{\hat N}.
\end{align}
Combining (\ref{AF-chi-odd}) and (\ref{AF-varphi-eom}) we can express
$\chi_n$ in terms of $\varphi_1(=\phi)$,
\begin{eqnarray}\label{chi-rel}
\chi_n &=& \sum_{l=n}^{\hat N}\frac{S_l^2}{S_{n-1}}\,\Box^{2l-n+1}\varphi_1
-\sum_{l=n}^{\hat N}\frac{S_l^2 Q_{l+1}}{S_{n-1}}\,\Box^{2l-n}\varphi_1,
\\
&=& \sum_{l=n}^{\hat N} \frac{S_l^2}{S_{n-1}}
\left( \Box - Q_{l+1} \right) \phi_{2l-n},\qquad (n=1,...,\hat N).
\label{phi-to-chi2}
\end{eqnarray}
See Appendix B for the detail.

In order to get the relations between the HD parameter $a_n$ and
the AF parameters $Q_n$ and $S_n$,
let us compare ${\cal L}_{{\rm HD}}^{(2\hat N +1)}$ in (\ref{HDL}) and
${\cal L}_{{\rm AF}}^{2\hat N+1}$ in (\ref{AF-odd2}).
We decompose ${\cal L}_{{\rm HD}}^{(2\hat N+1)}$ as
\begin{align}\label{HDL-odd2}
{\cal L}_{{\rm HD}}^{(2\hat N+1)} =
\frac12\sum_{n=1}^{\hat N+1} a_{2n-1}\phi\Box^{2n-1}\phi
-\frac12\sum_{n=1}^{\hat N+1} a_{2n-2} \phi\Box^{2n-2}\phi
+ {\cal L}_{{\rm int}}(\phi).
\end{align}
Comparing (\ref{AF-odd2}) and (\ref{HDL-odd2}), we obtain the relations,
\begin{align}\label{rel1}
&Q_1 = m^2,
\quad
Q_{n+1} = \frac{a_{2n}}{a_{2n+1}},
\nn \\
&S_0^2 = 1,
\quad
S_n^2 =  a_{2n+1}, \qquad (n=1,...\hat N).
\end{align}
Then ${\cal L}_{{\rm AF}}^{2\hat N+1}$ and
${\cal L}_{{\rm HD}}^{2\hat N+1}$ become equivalent.

Similarly to the $N=$ even case we can reproduce the field equation
for $\phi$.
Inserting the relation for $\chi_1$ in (\ref{chi-rel}) into
the first equation in (\ref{AF-varphi-eom}), we can get
the field equation for $\varphi_1$,
\begin{align}\label{AF-eom2}
(\Box - Q_1)\varphi_1 + \sum_{n=1}^{\hat N} S_n^2
(\Box-Q_{n+1}) \Box^{2n}\varphi_1
+  {\cal L}_{{\rm int}}^{'}(\varphi_1) = 0.
\end{align}
Using the relation (\ref{rel1}) this equation reproduces the field
equation \eqref{HD-eom} for the HD field $\phi(=\varphi_1)$.

Plugging the relations (\ref{rel1}) into (\ref{phi-to-variphi2}) and
(\ref{phi-to-chi2}), we can also express the AF fields
($\varphi_n,\, \chi_n$) in terms of $\phi_n=\Box^n\phi$ with the
coefficient $a_n$ in the HD field Lagrangian,
\begin{align}\label{HDtoAF2}
\varphi_m &= \sqrt{a_{2m-1}}\, \phi_{m-1}, \qquad (m=1,\cdots, \hat N+1),
\nn \\
\chi_n &= \sum_{l=n}^{\hat N}\left( \frac{a_{2l+1}}{\sqrt{a_{2n-1}}}\,
\phi_{2l-n+1} - \frac{a_{2l}}{\sqrt{a_{2n-1}}}\, \phi_{2l-n}\right)
\nn \\
&= \sum_{m=n}^{2\hat N-n+1}(-1)^{m+n-1}\frac{a_{m+n}}{\sqrt{a_{2n-1}}}\,
\phi_m, \qquad  (n=1,\cdots,\hat N) .
\end{align}
Now we construct the transformation matrix $U^{(o)}_{ij}$ from the HD field
to the AF field. Similarly to (\ref{HDtoAF3}),
we write down the transformation relation as
\begin{align}\label{HDtoAF4}
\xi_i = \sum_{j=1}^{2\hat N+1} U^{(o)}_{ij}\phi_{j-1},
\qquad (i,\,j=1,\cdots, 2\hat N+1),
\end{align}
where $\xi$ is a column matrix with $2\hat N+1$ components,
\begin{align}\label{defxio}
\xi_m \equiv \varphi_m,\quad (m=1,\cdots, \hat N+1), \qquad
\xi_{\hat N+n+1}\equiv \chi_n,\quad (n=1,\cdots, \hat N).
\end{align}
We read the transformation matrix $U^{(o)}_{ij}$ from (\ref{HDtoAF2}),
\begin{align}
U^{(o)}_{ij} =\left\{
\begin{array}{ll}
&\left.
\begin{array}{ll}
\sqrt{a_{2j-1}}\,\delta_{ij} \qquad\qquad\quad &~
(1\le j\le\hat N+1)\\
0 \qquad\qquad\quad &~ (\mbox{the others})
\end{array}
\hskip 1.7cm\right\} (1\le i \le\hat N+1) \\
&\left.
\begin{array}{ll}
(-1)^{i+j-\hat N-1}
\frac{a_{i+j-\hat N-2}}{\sqrt{a_{2i-2\hat N-3}}} &
(i-\hat N\le j\le 3\hat N-i+3) \\
0 & (\mbox{the others})
\end{array}
\right\} (\hat N+2\le i \le 2\hat N+1)
\end{array}\right..
\end{align}
Again, using the inverse matrix of $U^{(o)}_{ij}$, we can also express
the HD field $\phi_i$ in terms of the AF fields $\xi_i$,
\begin{align}
\phi_{i-1} = \sum_{j=1}^{2\hat N +1} {U^{(o)}}^{-1}_{ij}\xi_j.
\end{align}

We apply the AF Lagrangian (\ref{AFL-odd}) for the next-to-minimal
higher derivative theory ($N=3$ case). This case was first
constructed by Carone and Lebed in Ref.~\cite{Carone:2008iw}.
The $N=3$ HD Lagrangian with $a_2=1/M_1^2$ and $a_3=1/M_2^4$
is given by
\begin{align}\label{HD-3}
{\cal L}_{{\rm HD}}^{(3)} = \frac{1}{2}\phi\Box\phi
- \frac{1}{2M_1^2} \phi\Box^2\phi
+\frac{1}{2M_2^4}\phi\Box^3\phi
- \frac{1}{2}m^2\phi^2 +{\cal L}_{{\rm int}}(\phi).
\end{align}
We read the $N=3$ AF Lagrangian from (\ref{AFL-odd}) as
\begin{align}\label{AF3}
{\cal L}_{{\rm AF}}^{(3)} &= \frac12\phi(\Box - m^2)\phi
+ \frac12\tilde\phi\Big(\Box - \frac{M_2^4}{M_1^2}\Big)\tilde\phi
+\chi (\Box\phi - M_2^2\tilde\phi) + {\cal L}_{{\rm int}} (\phi),
\end{align}
where we redefined the fields as
$\varphi_1=\phi,\, \varphi_2 = \tilde\phi$, and $\chi_1 = \chi$, and
used the identifications given in (\ref{rel1}),
\begin{align}
&Q_1 = m^2,\quad Q_2 = \frac{a_2}{a_3} = \frac{M_2^4}{M_1^2},
\quad S_1^2 = \frac1{R_1^2}=a_3 = \frac1{M_2^4}.
\end{align}

The AF Lagrangian (\ref{AF3}) is slightly different from the AF Lagrangian
given in Ref.~\cite{Carone:2008iw}. There
is an additional $\chi\phi$-term in Ref.~\cite{Carone:2008iw}.
Due to this term the normalization
in front of the quadratic terms is different from ours. However,
if we identify the parameters in (\ref{HDL}) and (\ref{AF3}), such as
\begin{align}\label{para3}
&m^2 = m_1^2m_2^2m_3^2/\Lambda^4,
\nn \\
&a_2 = M_1^{-2} = (m_1^2 + m_2^2 + m_3^2)/\Lambda^4,
\nn \\
&a_3= M_2^{-4} = 1/\Lambda^4
\end{align}
with $\Lambda^4 = m_1^2m_2^2 + m_2^2m_3^2 + m_3^2m_1^2$, we can see
the AF Lagrangian (\ref{AF3}) reproduces the same LW form given in
Ref.~\cite{Carone:2008iw}, under some linear mappings.
Therefore, the AF Lagrangians (\ref{AFL-even}) and (\ref{AFL-odd})
that we constructed are not unique.
However, the resulting LW form is unique,
which will be studied in the next section.

\section{Generalized Lee-Wick Form}

In the previous section we showed that the HD Lagrangian can be
recast into the AF Lagrangian by trading the higher derivatives with
additional fields ($\varphi_{n>1}$) including auxiliary fields ($\chi_n$).
The $N$ physical poles in the HD-field propagator are
converted into $N$ degrees of freedom in the AF Lagrangian.
If we construct the LW
form by appropriate linear combinations of the AF fields,
as done in Refs.~\cite{Grinstein:2007mp,Carone:2008iw},
the resulting LW form is equivalent to the original HD Lagrangian
up to the quantum level.
In this section, we complete constructing the transformations among
the three forms of Lagrangian, the HD Lagrangian, the AF Lagrangian,
and the LW form, for arbitrary $N$.

\subsection{HD Lagrangian and LW form}\label{HDLW}

In this subsection, we reconstruct and extend the Pais-Uhlenbeck
formalism~\cite{Pais:1950za} which relates  the HD
Lagrangian and the LW form.
We explicitly express the LW field $\psi_n$ in terms
of the HD field ($\Box^n\phi$, $n=1,...,N$) and HD parameters
$m^2$ and $a_n$.

We consider the following LW form corresponding to the HD Lagrangian
with $N$ physical poles in the HD field propagator,
\begin{align}\label{LWL}
{\cal L}_{{\rm LW}}^{(N)} = \frac12\sum_{n=1}^{N} \kappa_n
\psi_n(\Box - \mu_n)\psi_n + {\cal L}_{{\rm int}}(\psi),
\end{align}
where $\kappa_n$ represents the overall sign $(\pm 1)$,
$\mu_n$ is the mass-squared parameter, and $\psi_n$ denotes the LW field.
The HD field $\phi$ can be expressed as a linear combination of $\psi_n$,
\begin{align}
\phi= \sum_{n=1}^N c_n\psi_n,
\nn
\end{align}
where $c_n$ is a real coefficient. We assume that there is no
degeneracy of the mass-squared, i.e.,
\begin{align}\label{mu-n}
\mu_1 <\mu_2 < \mu_3 < ... <\mu_N.
\end{align}
The equation of motion for $\psi_n$ is given by
\begin{align}\label{LW-eom}
\kappa_n(\Box - \mu_n)\psi_n + c_n{\cal L}_{{\rm int}}^{'}(\psi) =0,
\end{align}
where ${\cal L}_{{\rm int}}^{'}= d{\cal L}_{{\rm int}}/d\psi_n$.

We compare the two Lagrangians (\ref{HDL}) and (\ref{LWL}),
and obtain the field transformation in a form,
\begin{align}\label{anspsi}
\psi_i = \sum_{j=1}^N b_{ij}\phi_{j-1},
\end{align}
where $b_{ij}$ is the component of the mapping matrix
from the HD field to the LW field, and $\phi_i = \Box^i\phi$.
We can rewrite the HD Lagrangian \eqref{HDL} as
\begin{align}\label{HDLF}
{\cal L}^{(N)}_{{\rm HD}} = \frac12\phi F(\Box)\phi +
{\cal L}_{{\rm int}} (\phi),\qquad
F(\Box) = -\sum_{n=0}^N (-1)^n a_n \Box^n,
\end{align}
where $F(\Box)$ is the inverse of the HD-field propagator.
Considering that the HD Lagrangian \eqref{HDLF} is equivalent to the
LW form \eqref{LWL}, the HD-field propagator can be expressed in
a form associated with $N$-physical poles,
\begin{align}\label{F2}
F(\Box) = -a_0 \prod_{i=1}^{N}\left(1-\frac{\Box}{\mu_i}\right).
\end{align}
By comparing the expressions of $F(\Box)$ in Eqs. \eqref{HDLF}
and \eqref{F2}, one can get the relations
between the HD parameter $a_n$ and the LW parameter $\mu_n$,
\begin{align}
a_1 &= a_0\sum_{i=1}^N\frac{1}{\mu_i},
\nn \\
a_2 &= a_0\sum_{i<j}^N\frac{1}{\mu_i\mu_j},
\nn \\
a_3 &= a_0 \sum_{i<j<k}^N\frac{1}{\mu_i\mu_j\mu_k},
\nn \\
&\hskip 1.5cm \vdots
\hskip 2cm  .
\label{a0-1}
\end{align}
In our parametrization, we set
\begin{align}
a_1=1 \quad \Leftrightarrow \quad
a_0 (= m^2) = \left(\sum_{i=1}^N\frac{1}{\mu_i}\right)^{-1}.
\end{align}

Using the ``partial-fraction analysis", we can express the
propagator as\footnote{Our parameters are related with
Pais-Uhlenbeck parameters by $\mu_i=\omega_i^2$ and $\eta_i/\mu_i^2
= \eta_i^{{\rm PU}}$.}
\begin{align}\label{invF}
\frac{1}{F(\Box)} = -\frac{1}{a_0}\prod_{i=1}^N\left(
1-\frac{\Box}{\mu_i}\right)^{-1} =
\sum_{i=1}^N \frac{\eta_i}{\Box - \mu_i},
\end{align}
where
\begin{align}
\eta_i = \frac{\mu_i}{a_0}\prod_{j\ne i}^N
\left(1-\frac{\mu_i}{\mu_j}\right)^{-1}.
\end{align}
Using the relation~\eqref{invF},
the HD Lagrangian \eqref{HDLF} reproduces the LW form \eqref{LWL}
(up to field rescaling),
\begin{align}
\frac12\phi F(\Box)\phi &= \frac12 \phi \left[F(\Box)\right]^2
\sum_{i=1}^N\frac{\eta_i}{\Box-\mu_i}\phi
\nn \\
&= \frac12\sum_{i=1}\frac{\eta_i}{\mu_i^2}\phi \left[F(\Box)\right]^2
\frac{\mu_i^2}{\Box-\mu_i}\phi
\nn \\
&= \frac12 \sum_{i=1} a_0^2 \frac{\eta_i}{\mu_i^2} \phi
\left[\prod_{j\ne i}^N\Big(1-\frac{\Box}{\mu_j}\Big)\right]^2
(\Box-\mu_i)\phi
\nn \\
&= \frac12 \sum_{i=1}\kappa_i\psi_i (\Box-\mu_i)\psi_i,
\end{align}
from which we can get the transformation,
\begin{align}\label{psi1}
\psi_i  =
a_0\frac{\sqrt{|\eta_i|}}{\mu_i}\prod_{j\ne i}^N \left(
1-\frac{\Box}{\mu_i}\right)\phi= \sum_{j=1}^N b_{ij}\phi_{j-1}.
\end{align}
The elements of this transformation matrix are given by
\begin{align}
b_{i1} &= \frac{a_0}{\mu_i}\sqrt{|\eta_i|},
\nn \\
b_{i2}&= -\frac{a_0}{\mu_i}\sqrt{|\eta_i|} \sum_{j\ne i}^N (\mu_j)^{-1},
\nn \\
b_{i3}&= \frac{a_0}{\mu_i}\sqrt{|\eta_i|}
\sum_{j (\ne i) < k (\ne i)}^N (\mu_j\mu_k)^{-1},
\nn \\
&\hskip 1.5cm \vdots
\hskip 4cm  .
\label{bnm}
\end{align}

We can now obtain the inverse transformation from the LW field to 
the HD field,
\begin{align}\label{pbpsi}
\phi_{j-1} = \sum_{i=1}^N (b^{-1})_{ji} \psi_i .
\end{align}
Using the relation \eqref{invF}, we obtain for $j=1$,
\begin{align}
\phi &= \left( \sum_{i=1}^N \frac{\eta_i}{\Box-\mu_i} \right) F(\Box)\phi
= \sum_{i=1}^N a_0\frac{\eta_i}{\mu_i} \prod_{j\ne i}^N \left(
1-\frac{\Box}{\mu_j}\right)\phi
= \sum_{i=1}^N (b^{-1})_{1i}\psi_i,
\end{align}
where
\begin{align}
(b^{-1})_{1i} = \kappa_i \sqrt{|\eta_i|},
\end{align}
and $\kappa_i$ is the signature of $\eta_i$.
Then
\begin{align}
\phi_{j-1} &= \Box^{j-1} \phi = \sum_{i=1}^N \kappa_i \sqrt{|\eta_i|}
\Box^{j-1}\psi_i = \sum_{i=1}^N \kappa_i \sqrt{|\eta_i|}\mu_i^{j-1}\psi_i
\nn\\
&= \sum_{i=1}^N (b^{-1})_{ji}\psi_i,
\end{align}
therefore, we have
\begin{align}
(b^{-1})_{ji} = \mu_i^{j-1} (b^{-1})_{1i} = 
\mu_i^{j-1} \kappa_i \sqrt{|\eta_i|}.
\end{align}
There are simple algebraic sum-rules for $\eta_i$,
\begin{align}
\sum_{i=1}^N \frac{\eta_i}{\mu_i} = a_0,\qquad
\sum_{i=1}^N \eta_i\mu_i^n =0,\quad (n=0,1,\cdots,N-2).
\end{align}

Now we present several examples for $N=2,3$, and $4$.

\noindent \underline{\bf $N=2$ case:}
\\ \noindent
For simplicity, let $\mu_{ij}\equiv \mu_i-\mu_j$.
\\ \noindent
(i) The HD coefficient $a_n$ is completely determined by the LW-mass
parameter from Eq.~\eqref{a0-1},
\be \left(
  \begin{array}{c}
    a_0 \\
    a_2 \\
  \end{array}
\right) = \frac{1}{\mu_1+\mu_2} \left(
  \begin{array}{c}
    \mu_1\mu_2 \\
    1 \\
  \end{array}
\right). \ee

\noindent
(ii) The transformation matrix $b_{ij}$ is
obtained by Eq.~\eqref{bnm},
\be \left(
  \begin{array}{c}
    b_{ij} \\
  \end{array}
\right) =  \frac{1}{\sqrt{\mu_{21}(\mu_1+\mu_2)}} \left(
  \begin{array}{cc}
    \mu_2 & -1 \\
    \mu_1 & -1 \\
  \end{array}
\right). \ee

\vspace{12pt} \noindent \underline{\bf $N=3$ case:}
\\ \noindent
We follow the same process above.
\\
\noindent (i)
\be \left(
  \begin{array}{c}
    a_0 \\
    a_2 \\
    a_3 \\
  \end{array}
\right) = \frac{1}{\mu_1\mu_2+\mu_2\mu_3+\mu_3\mu_1} \left(
  \begin{array}{c}
    \mu_1\mu_2\mu_3 \\
    \mu_1+\mu_2+\mu_3 \\
    1 \\
  \end{array}
\right) \ee

\noindent (ii)
\be \left(
  \begin{array}{c}
    b_{ij} \\
  \end{array}
\right) = \frac{1}{\sqrt{\mu_1\mu_2+\mu_2\mu_3+\mu_3\mu_1}} \left(
  \begin{array}{ccc}
    \frac{\mu_2\mu_3}{\sqrt{\mu_{31}\mu_{21}}}
    & -\frac{\mu_2+\mu_3}{\sqrt{\mu_{31}\mu_{21}}}
    & \frac{1}{\sqrt{\mu_{31}\mu_{21}}} \\
    \frac{\mu_3\mu_1}{\sqrt{\mu_{32}\mu_{21}}}
    & -\frac{\mu_3+\mu_1}{\sqrt{\mu_{32}\mu_{21}}}
    & \frac{1}{\sqrt{\mu_{32}\mu_{21}}} \\
    \frac{\mu_1\mu_2}{\sqrt{\mu_{32}\mu_{31}}}
    & -\frac{\mu_1+\mu_2}{\sqrt{\mu_{32}\mu_{31}}}
    & \frac{1}{\sqrt{\mu_{32}\mu_{31}}} \\
  \end{array}
\right) \ee

\vspace{12pt} \noindent \underline{\bf $N=4$ case:}
\\ \noindent
For simplicity, let $\mu_A =
\mu_1\mu_2\mu_3+\mu_2\mu_3\mu_4+\mu_3\mu_4\mu_1+\mu_4\mu_1\mu_2$.

\noindent (i)
\be \left(
  \begin{array}{c}
    a_0 \\
    a_2 \\
    a_3 \\
    a_4 \\
  \end{array}
\right) = \frac{1}{\mu_A} \left(
  \begin{array}{c}
    \mu_1\mu_2\mu_3\mu_4 \\
    \mu_1\mu_2+\mu_2\mu_3+\mu_3\mu_4+\mu_4\mu_1+\mu_1\mu_3+\mu_2\mu_4 \\
    \mu_1+\mu_2+\mu_3+\mu_4 \\
    1 \\
  \end{array}
\right) \ee

\noindent (ii)
\be \left(
  \begin{array}{c}
    b_{ij} \\
  \end{array}
\right) = \frac{1}{\sqrt{\mu_A}} \left(
  \begin{array}{cccc}
    \frac{\mu_2\mu_3\mu_4}{\sqrt{\mu_{41}\mu_{31}\mu_{21}}}
    & \frac{-(\mu_2\mu_3+\mu_2\mu_4+\mu_3\mu_4)}{\sqrt{\mu_{41}\mu_{31}\mu_{21}}}
    & \frac{\mu_2+\mu_3+\mu_4}{\sqrt{\mu_{41}\mu_{31}\mu_{21}}}
    & \frac{-1}{\sqrt{\mu_{41}\mu_{31}\mu_{21}}}  \\
    \frac{\mu_3\mu_4\mu_1}{\sqrt{\mu_{42}\mu_{32}\mu_{21}}}
    & \frac{-(\mu_1\mu_3+\mu_1\mu_4+\mu_3\mu_4)}{\sqrt{\mu_{42}\mu_{32}\mu_{21}}}
    & \frac{\mu_3+\mu_4+\mu_1}{\sqrt{\mu_{42}\mu_{32}\mu_{21}}}
    & \frac{-1}{\sqrt{\mu_{42}\mu_{32}\mu_{21}}}  \\
    \frac{\mu_4\mu_1\mu_2}{\sqrt{\mu_{43}\mu_{32}\mu_{31}}}
    & \frac{-(\mu_1\mu_2+\mu_1\mu_4+\mu_2\mu_4)}{\sqrt{\mu_{43}\mu_{32}\mu_{31}}}
    & \frac{\mu_4+\mu_1+\mu_2}{\sqrt{\mu_{43}\mu_{32}\mu_{31}}}
    & \frac{-1}{\sqrt{\mu_{43}\mu_{32}\mu_{31}}}  \\
    \frac{\mu_1\mu_2\mu_3}{\sqrt{\mu_{43}\mu_{42}\mu_{41}}}
    & \frac{-(\mu_1\mu_2+\mu_1\mu_3+\mu_2\mu_3)}{\sqrt{\mu_{43}\mu_{42}\mu_{41}}}
    & \frac{\mu_1+\mu_2+\mu_3}{\sqrt{\mu_{43}\mu_{42}\mu_{41}}}
    &\frac{-1}{\sqrt{\mu_{43}\mu_{42}\mu_{41}}}  \\
  \end{array}
\right) \ee

\subsection{AF Lagrangian and LW form}
In this subsection, we construct the transformation matrix
from the LW field $\psi_n$ to the the auxiliary field $(\varphi_n,\chi_n)$.
Once $a_n$ is expressed by $\mu_n$ from Eq.~\eqref{a0-1},
the AF coefficients $Q_n$ and $R_n$ (so, $S_n$) can be expressed
by the LW parameter $\mu_n$
from Eqs.~\eqref{rel-even} and \eqref{rel1}.
Then the transformation matrix is completely expressed by $\mu_n$.
Using the inverse of this transformation matrix we can obtain the LW form
\eqref{LWL} from the AF Lagrangians \eqref{AFL-even} and
\eqref{AFL-odd}.

\subsubsection{$N=2\hat N$ case}

Plugging Eq.~\eqref{pbpsi} into Eq.~\eqref{HDtoAF3}, we find the mapping,
\begin{align}
\xi_i = \sum_{j=1}^{N} V^{(e)}_{ij}\psi_j, \qquad
(i,j =1,\cdots, N=2\hat N),
\end{align}
where $\xi_i$ was defined in (\ref{defxie}) and
\begin{align}
V^{(e)}_{ij} = \sum_{k=1}^{2\hat N} U^{(e)}_{ik} (b^{-1})_{kj}.
\end{align}

We present the examples for $N=2$ and $N=4$.

\noindent \underline{\bf $N=2$ case:}
\\ \noindent
\begin{align} \left(
  \begin{array}{c}
    \varphi_n \\
  \end{array}
\right) &=\frac{\mu_{21}}{\mu_2+\mu_1} \left(
  \begin{array}{cc}
    +1 & -1 \\
  \end{array}
\right) \left(
  \begin{array}{c}
    \psi_m \\
  \end{array}
\right)
\\
\left(
  \begin{array}{c}
    \chi_n \\
  \end{array}
\right) &=\left[ \frac{1}{\sqrt{\mu_{21}(\mu_2+\mu_1)}} \left(
  \begin{array}{cc}
    -\mu_1 & \mu_2 \\
  \end{array}
\right)+ \left(
  \begin{array}{cc}
    0 & 0 \\
  \end{array}
\right) \right]
\left(
  \begin{array}{c}
    \psi_m \\
  \end{array}
\right)
\end{align}

\vspace{12pt} \noindent \underline{\bf $N=4$ case:}
\\ \noindent
For simplicity, let $\mu_B = \mu_1+\mu_2+\mu_3+\mu_4$.
\begin{align} \left(
  \begin{array}{c}
    \varphi_n \\
  \end{array}
\right) &= \left(
  \begin{array}{cccc}
    \sqrt{\frac{\mu_A}{
    \mu_{41}\mu_{31}\mu_{21}}}
    & -\sqrt{\frac{\mu_A}{
    \mu_{42}\mu_{32}\mu_{21}}}
    & \sqrt{\frac{\mu_A}{
    \mu_{43}\mu_{32}\mu_{31}}}
    & -\sqrt{\frac{\mu_A}{
    \mu_{43}\mu_{42}\mu_{41}}} \\
    \mu_1\sqrt{\frac{\mu_B}{
    \mu_{41}\mu_{31}\mu_{21}}}
    & -\mu_2\sqrt{\frac{\mu_B}{
    \mu_{42}\mu_{32}\mu_{21}}}
    & \mu_3\sqrt{\frac{\mu_B}{
    \mu_{43}\mu_{32}\mu_{31}}}
    & -\mu_4\sqrt{\frac{\mu_B}{
    \mu_{43}\mu_{42}\mu_{41}}} \\
  \end{array}
\right) \left(
  \begin{array}{c}
    \psi_m \\
  \end{array}
\right) \\
 \left(
  \begin{array}{c}
    \chi_n \\
  \end{array}
\right) &= \left(
  \begin{array}{cccc}
    -\frac{\mu_1(\mu_2\mu_3+\mu_2\mu_4+\mu_3\mu_4)}{\sqrt{
    \mu_A\mu_{41}\mu_{31}\mu_{21}}}
    & \frac{\mu_2(\mu_1\mu_3+\mu_1\mu_4+\mu_3\mu_4)}{\sqrt{
    \mu_A\mu_{42}\mu_{32}\mu_{21}}}
    & -\frac{\mu_3(\mu_1\mu_2+\mu_1\mu_4+\mu_2\mu_4)}{\sqrt{
    \mu_A\mu_{43}\mu_{32}\mu_{31}}}
    & \frac{\mu_4(\mu_1\mu_2+\mu_1\mu_3+\mu_2\mu_3)}{\sqrt{
    \mu_A\mu_{43}\mu_{42}\mu_{41}}} \\
    -\frac{\mu_1^2}{\sqrt{
    \mu_B\mu_{41}\mu_{31}\mu_{21}}}
    & \frac{\mu_2^2}{\sqrt{
    \mu_B\mu_{42}\mu_{32}\mu_{21}}}
    & -\frac{\mu_3^2}{\sqrt{
    \mu_B\mu_{43}\mu_{32}\mu_{31}}}
    & \frac{\mu_4^2}{\sqrt{
    \mu_B\mu_{43}\mu_{42}\mu_{41}}} \\
  \end{array}
\right) \left(
  \begin{array}{c}
    \psi_m \\
  \end{array}
\right)
\end{align}

\subsubsection{$N=2\hat N+1$ case}

Plugging Eq.~\eqref{pbpsi} into Eq.~\eqref{HDtoAF4}, we find the mapping,
\begin{align}
\xi_i = \sum_{j=1}^{N} V^{(o)}_{ij}\psi_j, \qquad
(i,j =1,\cdots, N=2\hat N+1),
\end{align}
where $\xi_i$ was defined in (\ref{defxio}) and
\begin{align}
V^{(o)}_{ij} = \sum_{k=1}^{2\hat N +1} U^{(o)}_{ik} (b^{-1})_{kj}.
\end{align}

\vspace{12pt} \noindent \underline{\bf $N=3$ case:}
\\ \noindent
\begin{align} \left(
  \begin{array}{c}
    \varphi_n \\
  \end{array}
\right) &= \left(
  \begin{array}{ccc}
    \sqrt{\frac{\mu_1\mu_2+\mu_2\mu_3+\mu_3\mu_1}{\mu_{21}\mu_{31}}}
    & -\sqrt{\frac{\mu_1\mu_2+\mu_2\mu_3+\mu_3\mu_1}{\mu_{21}\mu_{32}}}
    & \sqrt{\frac{\mu_1\mu_2+\mu_2\mu_3+\mu_3\mu_1}{\mu_{31}\mu_{32}}} \\
    \frac{\mu_1}{\sqrt{\mu_{21}\mu_{31}}}
    & -\frac{\mu_2}{\sqrt{\mu_{21}\mu_{32}}}
    & \frac{\mu_3}{\sqrt{\mu_{31}\mu_{32}}} \\
  \end{array}
\right) \left(
  \begin{array}{c}
    \psi_m \\
  \end{array}
\right) \\
\left(
  \begin{array}{c}
    \chi_n \\
  \end{array}
\right) &=\frac{1}{\sqrt{\mu_1\mu_2+\mu_2\mu_3+\mu_3\mu_1}} \left(
  \begin{array}{ccc}
    -\frac{\mu_1(\mu_2+\mu_3)}{\sqrt{\mu_{21}\mu_{31}}}
    & \frac{\mu_2(\mu_3+\mu_1)}{\sqrt{\mu_{21}\mu_{32}}}
    & -\frac{\mu_3(\mu_1+\mu_2)}{\sqrt{\mu_{31}\mu_{32}}}
  \end{array}
\right) \left(
  \begin{array}{c}
    \psi_m \\
  \end{array}
\right)
\end{align}

\section{Conclusions}

In this work we considered a HD field theory
with $N$ physical poles
of the two-point function for a self-interacting real scalar field.
We generalized the Carone and Lebed's work~\cite{Carone:2008iw},
in which the AF Lagrangian
and the LW form for the $N=3$ HD Lagrangian were constructed.

Our work is summarized as follows;

\noindent
(i) In section 2, for an HD Lagrangian with arbitrary $N$ poles,
we obtained the corresponding AF Lagrangian.
We obtained the explicit transformation from the HD field to the AF field,
and the inverse transformation is also possibly obtained.

\noindent
(ii) In section 3.1, we found the transformation from the HD field
to the LW field,
and the inverse transformation was obtained
explicitly.

\noindent
(iii) In section 3.2, we showed that the corresponding LW form
can be constructed by the redefinitions of the AF field
using the transformation from the LW field to the AF field.
\noindent

We presented the results of our formulation for $N=2,3,$ and $4$.
However, in principle, one can obtain the results for $N>4$ from
our formulation.

There are several things to comment on the AF Lagrangian, which is one of our
main results. In the construction of the AF Lagrangian,
we split the HD Lagrangian into $N=$ even and $N=$ odd cases,
since they are qualitatively different.
For the $N=$ even case, we introduced a quadratic term for the
last component of auxiliary field ($\chi_{\hat N}^2$-term),
while the other auxiliary fields are linear.
For the $N=$ odd case, however, every auxiliary field is linear.
Since the AF fields are linear or quadratic in the Lagrangian,
the resulting equations of motion impose constraints which are exact
at the quantum level.
We could also obtain the LW form from the AF Lagrangian by linear mappings.
Therefore, the HD Lagrangian, the AF Lagrangian, and the LW form are equivalent
up to the quantum level also.

As we did in section \ref{HDLW}, we can directly find the mapping
matrix between the HD field and the LW field. In doing so, we could
use the ``partial-fraction analysis" since the numerators of the LW
propagators are numbers ($\eta_i$ in our case). By rescaling the
fields we can always absorb the numerical factors. For this reason,
we can actually prove that the equivalence between the HD lagrangian
and the LW lagrangian up to the quantum level, without introducing
the AF lagrangian.

When gauge fields are introduced in the theory, being different from
the scalar-field case, the numerators of the propagators contain the
gauge indices (for non-abelian gauge fields) as well as the
spacetime indices. Therefore, the partial-fraction analysis is not
useful for the gauge-field case. In order to prove the quantum
equivalence between the HD lagrangian and the LW lagrangian, we need
to introduce the AF lagrangian as an intermediate step. In
Ref.~\cite{Carone:2008iw}, Carone and Lebed applied the results of
the scalar field ($m_1=0$ case) to construct the AF lagrangian and
the LW lagrangian from the $N=3$ HD lagrangian with the gauge field.
When the extension is to be made for $N>3$ with gauge fields in the
future, the AF lagrangian formalism which we constructed for the
scalar field in this work will be very useful.

Another thing that we have to mention is that the AF Lagrangian is not
unique, while the LW form is unique for a given HD Lagrangian.
For example, the $N=3$ AF Lagrangian constructed in Ref.~\cite{Carone:2008iw}
is different from ours (\ref{AF3}). We can easily check this fact by
comparing the overall normalizations of the two AF Lagrangians.
However, the $N=3$ LW form in Ref.~\cite{Carone:2008iw} and ours
(see section \ref{HDLW}) are definitely identical.

Our $N$-field formulation can be applied to various interesting
physical situations, for example, $N$-field Lee-Wick phenomenology,
$N$-field cosmology in a relation to cosmological expansions, and etc.
From the top-down aspect, once a higher derivative scalar field
theory is developed, one can apply our formulation in order
to see its phenomenological consequences.
From the bottom-up aspect, one can deduce the theoretical set-up
from the $N$-field interpretation of the phenomena.

\subsection*{Acknowledgements}
We are grateful to Hiroaki Nakajima, Corneliu Sochichiu, and Driba Tolla
for helpful discussions, and
the Asia Pacific Center for Theoretical Physics (APCTP) for its hospitality
during the workshop  ``APCTP Focus Program on Current Trends
in String Field Theory'', where a  part of this work was done.
This work was supported by the Korea Research Foundation (KRF) grant
funded by the Korea government (MEST) No. 2009-0070303 (I.Y.),
and No. 2009-0073775 (O.K.).

\appendix

\section{Calculation of (\ref{chi-even})}
Inserting the relations in Eq.~\eqref{AF-chi-even} into those in
Eq.~\eqref{AF-varphi-even}, we obtain
\begin{align}
\chi_{\hat N-1} &= \frac1{R_{\hat N-1}}(\Box - Q_{\hat N})\varphi_{\hat N}
+\frac1{R_{\hat N-1}}\Box\chi_{\hat N}
=\frac{S_{\hat N-1}}{R_{\hat N-1}} (\Box - Q_{\hat N})\Box^{\hat N-1}\varphi_1
-\frac{S_{\hat N}Q_{\hat N+1}}{R_{\hat N-1}R_{\hat N}}\Box^{\hat N+1}\varphi_1,
\nn \\
\chi_{\hat N-2} &= \frac1{R_{\hat N-2}}(\Box - Q_{\hat N-1})\varphi_{\hat N-1}
+\frac1{R_{\hat N-2}}\Box\chi_{\hat N-1}
\nn \\
&=\frac{S_{\hat N-2}}{R_{\hat N-2}} (\Box - Q_{\hat N-1})
\Box^{\hat N-2}\varphi_1 + \frac{S_{\hat N-1}}{R_{\hat N-2}R_{\hat N-1}}
(\Box - Q_{\hat N}) \Box^{\hat N}\varphi_1
-\frac{S_{\hat N}Q_{\hat N+1}}{R_{\hat N-2}R_{\hat N-1}R_{\hat N}}
\Box^{\hat N+2}\varphi_1,
\nn \\
&\hskip 2cm \vdots
\nn \\
\chi_1 &= \frac1{R_{1}}(\Box - Q_{2})\varphi_{2}
+\frac1{R_{1}}\Box\chi_{2}
\nn \\
&=\sum_{n=1}^{\hat N-1} S_n^2(\Box - Q_{n+1})
\Box^{2n-1}\varphi_1-S_{\hat N}^2Q_{\hat N+1}\Box^{2\hat N-1}\varphi_1.
\end{align}

\section{Calculation of (\ref{chi-rel})}
Inserting the relations in Eq.~\eqref{AF-chi-odd} into those in
Eq.~\eqref{AF-varphi-eom}, we obtain
\begin{align}
\chi_{\hat N} &= \frac1{R_{\hat N}}(\Box - Q_{\hat N+1})\varphi_{\hat N+1}
=\frac{S_{\hat N}}{R_{\hat N}} (\Box - Q_{\hat N+1})\Box^{\hat N}\varphi_1,
\nn \\
\chi_{\hat N-1} &= \frac{S_{\hat N-1}}{R_{\hat N-1}} (\Box - Q_{\hat N})
\Box^{\hat N-1}\varphi_1 + \frac{S_{\hat N}}{R_{\hat N-1}R_{\hat N}}
(\Box - Q_{\hat N+1}) \Box^{\hat N+1}\varphi_1,
\nn \\
&\hskip 2cm \vdots
\nn \\
\chi_1 &= \sum_{n=1}^{\hat N} S_n^2(\Box - Q_{n+1})
\Box^{2n-1}\varphi_1.
\end{align}

\end{document}